\documentclass{article}
          \let\footnote\savefootnote
          \let\footnotetext\savefootnotetext
          
\begin{document}
          \title{The Representation of Numbers by States in Quantum Mechanics}
          \author{Paul Benioff \\
           Physics Division, Argonne National Laboratory \\
           Argonne, IL 60439, \\
           e-mail: pbenioff@anl.gov}

        \maketitle

\begin{abstract}
The representation of numbers by tensor product states of
composite quantum systems is examined. Consideration is limited
to $k-ary$ representations of length $L$ and arithmetic $\bmod
k^{L}$. An abstract representation on an $L$ fold tensor product
Hilbert space ${\mathcal H}^{arith}$ of number states and
operators for the basic arithmetic operations is described.
Unitary maps onto a physical parameter based  tensor product
space ${\mathcal H}^{phy}$ are defined and the relations between
these two spaces and the dependence of algorithm dynamics on the
unitary maps is discussed.  The important condition of efficient
implementation by physically realizable Hamiltonians of the basic
arithmetic operations is also discussed.
\end{abstract}

\section{INTRODUCTION}

 The  representation of numbers by states of physical systems is basic
and widespread in science.  However, this representation is
assumed and used implicitly, with little effort devoted to
exactly what assumptions and conditions are implied.  Here this
question is examined for the nonnegative integers. Consideration
will be limited to $k-ary$ representations of length $L$ of
numbers  by tensor product states and to arithmetic modulo
$k^{L}$.

Based on the universality of quantum mechanics, all physical
systems of interest are quantum systems described by pure or
mixed quantum states. This is the case whether the systems are
microscopic, as is the case for quantum computers, or
macroscopic, as is the case for all presently existing computers.
Microscopic systems are those for which the ratio
$t_{dec}/t_{sw}>>1$ where $t_{dec}$ and $t_{sw}$ are the
decoherence and switching times \cite{DiVincenzo}. In this case
the system remains isolated from the environment for a time
duration of many switching steps. If $t_{dec}/t_{sw}<1$ then the
system is macroscopic, and environmental interactions stabilize
some system states  (the pointer states) \cite{Zurek1} for a time
duration of many switching steps. The emphasis here is on
microscopic systems although much of the material also holds for
macroscopic systems.

One route to the exact meaning of the representation of numbers
by states of physical systems begins with pure mathematics.  A
nonempty set is a set of numbers if and only if it satisfies
(i.e. is a model of) the axioms of number theory or arithmetic
\cite{Shoenfield}. Here these axioms need to be modified by
inclusion of relevant axioms for a commutative ring with identity
as these axioms are satisfied by modular arithmetic
\cite{Adamson}. Details of the axioms are not important here.
However, necessary conditions for a nonempty set to be a model
include the existence of functions or operators with the
properties of basic arithmetic operations, the successor $S$ (or
$+1$), $+$, and $\times$, given by the axioms. The ordering
relation and the induction schema will not be discussed here.

If the axioms are consistent then there are many mathematical
models of the axioms. Included are models containing tensor
product states and unitary operators on a product Hilbert space
of states. A  model  on an abstract product Hilbert space
${\mathcal H}^{arith}$ is described in Section \ref{MBAR}.

The connection to physics is made in two steps. First a Hilbert
space ${\mathcal H}^{phy}$ based on two sets of physical
parameters is described.  Models, based on ${\mathcal H}^{phy}$,
of the axioms are described in Section \ref{PPQM}.  These are
generated by unitary operators from ${\mathcal H}^{arith}$ to
${\mathcal H}^{phy}$.

The second step is the requirement that there exist physical
models of the axioms in which the basic arithmetic operations are
efficiently implementable. The widespread existence of computers
shows that this existence requirement is satisfied, at least for
macroscopic systems.

This requirement is discussed in Section \ref{EIAO} and applied to
models based on ${\mathcal H}^{phy}$. Due to space limitations the
discussion in this and other sections is brief. Details and
proofs are provided elsewhere \cite{BenRNQM}.

\section{Models Based on ${\mathcal H}^{arith}$} \label{MBAR}
Let ${\mathcal H}^{arith}=\otimes_{j=1}^{L}{\mathcal H}_{j}$ where
${\mathcal H}_{j}$ is a $k$ dimensional Hilbert space spanned by
states $|h,j\rangle$ where $0\leq h\leq k-1$ and $j$ is fixed.
States in the corresponding basis spanning ${\mathcal H}^{arith}$
have the form $|\underline{s}\rangle =\otimes_{j=1}^{L}
|\underline{s}(j),j\rangle$ where $\underline{s}$ is any function
from $1,\cdots L$ to $0,\cdots ,k-1$. The value of $j$
distinguishes the component states (or qubytes) and $h$ ranges
over the $k$ possible values of the states for each component.

The reason $j$ is part of the state and not a subscript, as in
$\otimes_{j=1}^{L}|\underline{s}(j)\rangle_{j}$, is that the
action of operators to be defined depends on the value of $j$.
Expression of this dependence is not possible if $j$ appears as a
subscript and not between $|$ and $\rangle$.

Definitions of $S,\;+,\; \times$ are required by the axioms. The
efficient implementation requirement necessitates the definition
of  $L$ different successor operators, $V^{+1}_{j}$, for
$j=1,\cdots ,L$. These operators are defined to correspond to the
addition of $k^{j-1} \bmod k^{L}$ where $V^{+1}_{1}$ corresponds
to $S$.

To define the $V^{+1}_{j}$ let $u_{j}$ be a cyclic shift of
period $k$ that acts on the states $|h,j\rangle$ according to
$u_{j}|h,j\rangle=|h +1\bmod k,j\rangle$. $u_{j}$ is the identity
on all states $|m,j^{\prime}\rangle$ where $j^{\prime}\neq j$.
Define $V^{+1}_{j}$ by
\begin{eqnarray}
V^{+1}_{j} & = & \sum_{n=j}^{L}u_{n}P_{(\neq k-1),n} \prod_{\ell
=j}^{n-1}u_{\ell}P_{(k-1),\ell}
\nonumber \\
& & \mbox{} + \prod_{\ell = j}^{L} u_{\ell}P_{(k-1),\ell}
\label{p1expl}
\end{eqnarray}
Here $P_{(k-1),j} =|k-1,j\rangle\langle k-1,j| \otimes 1_{\neq
j}$ is the projection operator for finding  the $j$ component
state $|k-1,j\rangle$ and the other components in any state.
$P_{m,j}$ and $u_{j}$ satisfy the commutation relation
$u_{j}P_{m,j}=P_{m+1,j}u_{j} \bmod k$ for $m=0,\cdots ,k-1$. Also
$P_{(\neq k-1),j} = 1-P_{(k-1),j}$. In this equation the
unordered product is used because for any $p,q$, $u_{m}P_{p,m}$
commutes with $u_{n}P_{q,n}$ for $m\neq n$. Also for $n=j$ the
product factor with $j\leq \ell \leq n-1$ equals $1$.

The operator $+$ is defined on ${\mathcal
H}^{arith}\otimes{\mathcal H}^{arith}$ by $+|\underline{s}\rangle
|\underline{w}\rangle=|\underline{s}\rangle
|\underline{s+w}\rangle$ where
\begin{equation}
|\underline{s+w}\rangle = V^{+ \underline{s}(L)}_{L} V^{+
\underline{s}(L-1)}_{L-1} ,\cdots ,V^{+
\underline{s}(1)}_{1}|\underline{w}\rangle \label{+def}
\end{equation}
and $V^{+ \underline{s}(j)}_{j}=(V^{+1}_{j})^{\underline{s}(j)}$.

Note that for pairs of product states, which are first introduced
here, the domains of the functions $ \underline{s}$ and $
\underline{w}$ must be different.  That is $|\underline{s}\rangle
|\underline{w}\rangle=|\underline{s \ast w}\rangle$ where $\ast$
denotes concatenation and $\underline{s \ast w}$ is a function
from $1,\cdots ,2L$ to $0,\cdots ,k-1$. This  follows from the
requirement that all components of $|\underline{w}\rangle$ must
be distinguished from those in $|\underline{s}\rangle$.

There are some basic properties the operators $V^{+1}_{j}$ must
have: they are  cyclic shifts on ${\mathcal H}^{arith}$ and they
satisfy
\begin{equation}
V^{+1}_{j+1} = (V^{+1}_{j})^{k} \mbox{ for $1\leq j<L$:}\;\;\;
(V^{+1}_{L})^{k}=1 \label{logeff} \end{equation} This shows the
exponential dependence on $j$ and the need for separate
definitions and efficient implementation of each of these
operators. Also both the $V^{+1}_{j}$ and  $+$ are unitary. Proofs
of these and other properties and a definition of $\times$ are
given elsewhere \cite{BenRNQM}.

The proof that the operators $V^{+1}_{j},\; +,\; \times$ and
states of the form $|\underline{s}\rangle$ in ${\mathcal
H}^{arith}$ are a model of modular arithmetic consists in showing
that the appropriate axioms are satisfied.  Some details of this
are given in \cite{BenRNQM}. Note that ${\mathcal H}^{arith}$ is
also a model of the Hilbert space axioms.

\section{Models Based on ${\mathcal H}^{phy}$} \label{PPQM}
The model of modular arithmetic described so far is abstract.  No
connection to quantum physics is provided.  To remedy this one
needs to describe models based on physical parameters.  Let $A$
and $B$ be sets of $L$ and $k$ physical parameters for physical
observables $\hat{A},\; \hat{B}$ for an $L$ component quantum
system. The parameters in $A$  distinguish the $L$ component
systems from one another and the parameters of $B$ refer to $k$
different internal physical states of the component systems.
Examples of $A$ include a set of locations in space or a set of
excitation energies, as is used in NMR quantum computers
\cite{Gershenfeld}. Examples of $B$ include  spin projections
along an axis or energy levels of a particle in a potential well.

The physical parameter based Hilbert space ${\mathcal
H}^{phy}=\otimes_{a\epsilon A}{\mathcal H}_{a}$ where ${\mathcal
H}_{a}$ is spanned by states of the form $|b,a\rangle$ with
$b\epsilon B$ and a fixed. ${\mathcal H}^{phy}$ is spanned by
states $|\underline{t}\rangle=\otimes_{a\epsilon
A}|\underline{t}(a),a\rangle$ where $ \underline{t}$ is a
function from $A$ to $B$.

The presence of the $a$ component in the state
$|\underline{t}(a), a\rangle$ property in the state is essential
in that the state of the composite quantum system contains all the
quantum information available to any physical process or
algorithm.  It is used by  algorithms to distinguish the
different components or qubytes. This is especially the case for
any algorithm whose dynamics is described by a Hamiltonian that is
selfadjoint and time independent. This is an example of
Landauer's dictum "Information is Physical" \cite{Landauer}.

The goal here is for physical parameter states such as
$|\underline{t}\rangle$ to represent numbers.  However, it is
clear that the product states $|\underline{t}\rangle=
\otimes_{a\epsilon A} |\underline{t}(a),a\rangle$ do not
represent numbers.  The reason is that there is no association
between the labels $a$ and powers of $k$; also there is no
association between the range set $B$ of $\underline{t}$ and
numbers.

This can be remedied by use of unitary maps from ${\mathcal
H}^{arith}$ to ${\mathcal H}^{phy}$ that preserve the tensor
product structure. Let $g$ and $d$ be  bijections (one-one onto)
maps from $1,\cdots ,L$ to $A$ and from $0,\cdots ,k-1$ to $B$.
For each $g,d$, and $j$, $w_{g,d,j}$ is a unitary operator that
maps states $|h,j\rangle$ in ${\mathcal H}_{j}$ to states in
${\mathcal H}_{g(j)}$ according to $w_{g,d,j}|h,j\rangle =
|d(h),g(j)\rangle$. This induces a unitary operator
$W_{g,d}=\otimes_{j=1}^{L}w_{g,d,j}$ from ${\mathcal H}^{arith}$
to ${\mathcal H}^{phy}$ where
\begin{eqnarray}
W_{g,d}|\underline{s}\rangle = & \otimes_{j=1}^{L}w_{g,d,j}
|\underline{s}(j),j\rangle \nonumber \\ \mbox{} = &
\otimes_{j=1}^{L} |d(\underline{s}(j)),g(j)\rangle
=|\underline{s}_{g}^{d}\rangle. \label{wgddef}
\end{eqnarray}
Here $|\underline{s}_{g}^{d}\rangle$ is the physical parameter
based state in ${\mathcal H}^{phy}$ that corresponds, under
$W_{g,d}$ to the number state $|\underline{s}\rangle$ in
${\mathcal H}^{arith}$.

Conversely the adjoint operator $W^{\dagger}_{g,d}$  relates
physical parameter states in ${\mathcal H}^{phy}$ to number
states in ${\mathcal H}^{arith}$:
\begin{eqnarray}
W^{\dagger}_{g,d}|\underline{t}\rangle = & \otimes_{a\epsilon
A}w_{g^{-1},d^{-1},a} |\underline{t}(a),a\rangle \nonumber \\
\mbox{} = & \otimes_{a\epsilon A}
|d^{-1}(\underline{t}(a)),g^{-1}(a)\rangle
=|\underline{t}_{g^{-1}}^{d^{-1}}\rangle. \label{wdaggddef}
\end{eqnarray}
Here $|\underline{t}_{g^{-1}}^{d^{-1}}\rangle$ is the number state
in ${\mathcal H}^{arith}$ that corresponds to the physical state
$|\underline{t}\rangle$. Note that
$W^{\dagger}_{g,d}=W_{g^{-1},d^{-1}}$ where $g^{-1},d^{-1}$ are
the inverses of $g$ and $d$, and $w_{g^{-1},d^{-1},a} =
w^{\dagger}_{g,d,g^{-1}(a)}$.

The operators $W_{g,d}$ also induce representations of the
$V^{+1}_{j},\; +,$ and $\times$ operators on the physical
parameter states. For the $V^{+1}_{j}$ one defines
$V^{d,+1}_{g,j}$ by
\begin{equation}
V^{d,+1}_{g,j} = W_{g,d}V^{+1}_{j}W^{\dagger}_{g,d}.
\end{equation}
An equivalent definition can be obtained from Eq. \ref{p1expl} by
replacing $u_{\ell}$ by
$u^{d}_{g(\ell)}=w_{g,d,\ell}u_{\ell}w^{\dagger}_{g,d,\ell}$ and
$P_{k-1,\ell}$ by  $P_{d(k-1),g(\ell)} = w_{g,d,\ell}P_{k
-1,\ell}w^{\dagger}_{g,d,\ell}$.

In a similar fashion the $W_{g,d}$ are used to define $+_{g,d}$
acting on the physical parameter states in ${\mathcal
H}^{phy}\otimes {\mathcal H}^{phy}$.  The operator $+_{g,d}$ is
defined in terms of the operator $+$ on ${\mathcal
H}^{arith}\otimes {\mathcal H}^{arith}$ by
$+_{g,d}=(W_{g,d}\otimes W_{g,d})+(W^{\dagger}_{g,d}\otimes
W^{\dagger}_{g,d})$. $\times_{g,d}$ is defined similarly.

There are a large number of tensor product preserving unitary
maps from ${\mathcal H}^{arith}$ to ${\mathcal H}^{phy}$. Each of
these induces a model for the axioms of modular arithmetic on
${\mathcal H}^{phy}$. There are $L!k!$ of these maps restricted to
the form given by  Eq. \ref{wgddef} for $W_{g,d}$ as there are
$L!$ bijections $g$ and $k!$ bijections $d$. Thus there is no
unique correspondence between number states
$|\underline{s}\rangle$ in ${\mathcal H}^{arith}$ and physical
parameter states $|\underline{t}\rangle$ in ${\mathcal
H}^{phy}$.  In general the physical state
$|\underline{s}^{d}_{g}\rangle$ corresponding to the number state
$|\underline{s}\rangle$ depends on $g$ and $d$. Conversely, by
Eq. \ref{wdaggddef}, the number state
$|\underline{t}^{d^{-1}}_{g^{-1}}\rangle$ corresponding to the
physical state $|\underline{t}\rangle$ in ${\mathcal H}^{phy}$
depends on $g$ and $d$.

The question arises regarding the dependence of the dynamics of a
quantum algorithm on the $W_{g,d}$.  Some algorithms are
independent of these maps; others are not. In general, since the
the dynamics of any algorithm is physical, it must be described on
${\mathcal H}^{phy}$. If the algorithm can also be defined on
${\mathcal H}^{phy}$, then the dynamics is independent of these
maps. Grover's Algorithm \cite{Grover} is an example of this type
of algorithm as it can be described and implemented on states in
${\mathcal H}^{phy}$ that are linear superpositions of
$|\underline{t}\rangle$. No reference to states in ${\mathcal
H}^{arith}$ is needed.

However, arithmetic algorithms  must be defined on ${\mathcal
H}^{arith}$. For these the dynamics does depend on these maps.
Shor's Algorithm \cite{Shor} is an example of this as it
describes the computation of a numerical function
$f_{m}(x)=m^{x}\bmod M$ where $M$ and $m$ are relatively prime.

So far the discussion has been limited to models of modular
arithmetic on ${\mathcal H}^{arith}$ and on ${\mathcal H}^{phy}$.
However the full connection to physics has not yet been
established. This is given by the condition of efficient
implementation of the basic arithmetic operations.

\section{Efficient Implementability of the Basic Arithmetic Operations}
\label{EIAO} The meaning of this important requirement is that it
must be possible to physically implement the basic operations and
that the implementation must be efficient. That is, for the
$V^{d,+1}_{g,j}$, for each $j$ there must exist a physically
realizable Hamiltonian $H^{d}_{g,j}$ and a time $t_{j}$ such that
\begin{equation}
e^{-iH^{d}_{g,j}t_{j}}=V^{d,+1}_{g,j}.  \label{Hpr}
\end{equation}
This definition is quite general in that the Hamiltonian can depend on $j$.  However
for many systems and dynamics a Hamiltonian $H^{d}_{g}$ that implements
$V^{d,+1}_{g,j}$ and is independent of $j$ is realizable.

The requirement of efficiency means that both the space-time and
the thermodynamic resources required to physically implement the
operations must be at most polynomial in $L,k$. This condition
excludes $k=1$ or unary representations as all arithmetic
operations are exponentially hard in this case. Large values of
$k$ are also excluded as distinguishing among a large set of
symbols and carrying out simple arithmetic operations becomes
thermodynamically expensive. Also there are physical limitations
on the amount of information that can be reliably stored and
distinguished per unit space time volume \cite{Lloyd}.

Thermodynamic resources are needed to protect the system from errors resulting from
operation in a noisy environment. Microscopic systems also need protection from
decoherence \cite{Zurek}. Methods include the use of quantum error correction codes
\cite{QCerrors},  EPR pairs \cite{Bennett}, and decoherence free subspaces
\cite{Lidar}. Protection of macroscopic systems is less difficult since one takes
advantage of  decoherence to give stabilized "pointer" \cite{Zurek1} states that
represent numbers in a macroscopic computer.

The reason for separate definitions of the  $V^{d,+1}_{g,j}$  for each $j$ is that the
requirement means that each of these operators for $j=1,\cdots ,L$ must be efficiently
implemented. If the $V^{d,+1}_{g,j}$ were defined in terms of iterations of
$V^{d,+1}_{g,1}$,  then implementation of $V^{d,+1}_{g,j}$ would require $k^{j-1}$
iterations of $V^{d,+1}_{g,1}$.  This is not efficient even if $V^{d,+1}_{g,1}$ can be
efficiently implemented.

Since the $V^{d,+1}_{g,j}, +_{g,d}$, and $\times_{g,d}$ operators
are many system nonlocal operators, many dynamical steps  would be
needed to implement these operators by a realizable two particle
local Hamiltonian, Eq. \ref{Hpr}.  As is well known, there are
many methods of efficiently implementing these operators,  at
least in macroscopic computers. For the $V^{d,+1}_{g,j}$ methods
include moving the procedure for efficiently implementing
$V^{d,+1}_{g,1}$ along the path $g$ in $A$ to the site $g(j)$.
For $+_{g,d}$, methods are based on iterations of the
$V^{d,+1}_{g,j}$,  Eq. \ref{+def}.

The thermodynamic resources required to physically implement the $V^{d,+1}_{g,j}$ and
other arithmetic operations also depend on the path $g$. In general paths are chosen
that respect the topological or neighborhood properties of $A$ as this reduces the
resources required.  But in principle any path is possible as the resource dependence
on the path choice is not exponential.

\section{Discussion}
The importance of the efficient implementability condition must
be emphasized. Besides excluding $k=1$ and large values of $k$,
it excludes most unitary maps from ${\mathcal H}^{arith}$ to
${\mathcal H}^{phy}$.  To see this one notes that for any unitary
$U$, tensor product preserving or not, if the states
$|\underline{s}\rangle$ and operators $V^{+1}_{j},\; +,\; \times$
satisfy the axioms of modular arithmetic, so do the states
$U|\underline{s}\rangle$ and the operators
$UV^{+1}_{j}U^{\dagger},\; (U\otimes U)+(U^{\dagger}\otimes
U^{\dagger})$ and an operator for $\times$. However, for most $U$,
these operators are not efficiently implementable. Also the states
$U|\underline{s}\rangle$ may not be efficiently preparable. This
is the main reason the $U$ were taken to have the form $W_{g,d}$.

Much remains to be done. Future work includes  dropping the
modulo limitation and considering other types of numbers. The use
of annihilation and creation operators to represent states needs
examination. Also the exact meaning of physical realizability
needs to be clarified.

\section*{acknowledgments}
This work is supported by the U.S.
Department of Energy, Nuclear Physics Division, under contract
W-31-109-ENG-38.

           \end{document}